\documentclass[12pt,preprint]{aastex}

\shorttitle{Supernovae associated with off-axis GRBs}
\shortauthors{Waxman}

\begin{document}

\title{The nature of GRB980425 and the search for off-axis GRB signatures in nearby type Ib/c supernovae emission}

\author{Eli Waxman\altaffilmark{1}}
\altaffiltext{1}{Physics Faculty, Weizmann Institute of Science,
Rehovot 76100, Israel}

\begin{abstract}

The identification of type Ib/c supernovae as GRB progenitors is motivated by the association of GRB980425 with SN1998bw and of GRB030329 with SN2003dh. While the $\gamma$-ray luminosity of GRB030329 was typical to cosmological GRBs, the luminosity of the nearby (40~Mpc) GRB980425 was $\sim5$ orders of magnitude lower. The large luminosity difference is commonly explained by hypothesizing that either SNe Ib/c produce two different classes of GRBs, or that GRB980425 was a typical cosmological GRB jet viewed off-axis. In the latter scenario, strong radio emission, $L_\nu\sim10^{30}\nu_{10\rm GHz}^{-1/2}{\rm erg/s\,Hz}$, is expected at $\sim1$~yr delay due to jet deceleration to sub-relativistic speed, as observed from GRB970508. The radio luminosity of SN1998bw was 3 orders of magnitude lower than this value. We show that the low radio flux may be consistent with the off-axis jet interpretation, if the density of the wind surrounding the progenitor is lower than typically expected, $\dot{m}\equiv(\dot{M}/10^{-5}M_\odot{\rm yr}^{-1})/(v_{\rm w}/10^3{\rm km\,s^{-1}})\simeq0.1$ instead of $\dot{m}\gtrsim1$. The lower value of $\dot{m}$ is consistent with the observed radio emission from the supernova shock driven into the wind. This interpretation predicts transition to sub-relativistic expansion at $\sim10$~yr delay, with current $\approx1$~mJy 10~GHz flux and $m_V\approx23$ optical flux, and with $\approx10$~mas angular source size. It also implies that in order to search for the signature of off-axis GRBs associated with nearby Ib/c supernovae, follow up observations should be carried on a multi-yr time scale.

\end{abstract}


\section{Introduction}

The association of GRBs with type Ib/c supernovae is motivated by the temporal and angular coincidence of GRB980425 and SN1998bw \cite{Galama98b}, and by the identification of a SN1998bw-like spectrum in the optical afterglow of GRB030329 \cite{Stanek03,Hjorth03} (See, however, Katz 1994, who suggests that a SN like emission may result from the impact of the relativistic GRB debris on a nearby dense gas cloud). The $\gamma$-ray luminosity $L_{\gamma,\rm Iso.}\simeq10^{51}{\rm erg/s}$ of GRB030329, inferred from the redshift z=0.1685 of its host galaxy, lies within the range of typical cosmological GRBs, $L_{\gamma,\rm Iso.}\simeq10^{52\pm1}{\rm erg/s}$
(e.g. \cite{Schmidt01}). The subscript "Iso." indicates luminosity derived assuming isotropic emission. The association of GRB980425 with SN1998bw sets the distance to this burst to 38~Mpc (for $H_0=65{\rm km/s\,Mpc}$), implying that its luminosity is nearly 5 orders of magnitude lower than that typical for cosmological GRBs \cite{Pian00}.
 
Two hypotheses are commonly discussed, that may account for the orders of magnitude difference in luminosity. First, it may be that SNe Ib/c produce two different classes of GRBs, with two different characteristic luminosities. It is now commonly believed that long duration, $T>2$~s, cosmological, $L_{\gamma,\rm Iso.}\simeq10^{52}{\rm erg/s}$, GRBs are produced by the collapse of SN Ib/c progenitor stars. It is assumed that the stellar core collapses to a black-hole, which accretes mass over a long period, $\sim T$, driving a relativistic jet that penetrates the mantle/envelope and then produces the observed GRB \cite{Woosley93,Pac98,MacFadyen99}. This scenario is supported by the association of GRB030329 with SN2003dh, and by additional evidence for optical supernovae emission in several GRB afterglows \cite{Bloom03}. The origin of a second, low-luminosity, class is unknown. It may be, e.g., due to supernova shock break-out \cite{Colgate68,WoosleyWeaver86,MatznerMcKee99}. The small radius and high density of carbon/helium SN Ib/c progenitors may allow acceleration of the shock to mildly relativistic speed as it propagates through the steep density gradient near the stellar surface. It is not clear, however, that the energy transferred to mildly relativistic material is sufficient to account for the $\gamma$-ray emission. 
 
A second possibility is that GRB980425 was a typical, cosmological GRB jet viewed off-axis \cite{Nakamura98,Eichler99,Woosley99}\\ \cite{Granot02,Yamazaki03}. Due to the relativistic expansion of jet plasma, with Lorentz factor $\Gamma\gtrsim100$ during $\gamma$-ray emission \cite{Krolik,Baring}, $\gamma$-rays are concentrated into a cone of opening angle $\sim1/\Gamma$ around the expansion direction. Thus, if the jet is viewed from a direction making an angle larger than $\theta_j$+few~$\times1/\Gamma$ with the jet axis, where $\theta_j$ is the jet opening angle, the $\gamma$-ray flux may be strongly suppressed. In this scenario, strong radio emission, $L_\nu\sim10^{30}\nu_{10\rm GHz}^{-1/2}{\rm erg/s\,Hz}$, is expected at $\sim1$~yr delay \cite{FWK00,LnW00} as the jet decelerates to sub-relativistic speed and its emission becomes nearly isotropic (Perna \& Loeb (1998) have suggested that radio emission from the bow shock surrounding the jet may be detected on shorter time scale). Radio emission associated with transition to sub-relativistic expansion has been observed for GRB970508 (Frail, Waxman \& Kulkarni 2000, hereafter FWK00), for which the transition was inferred to occur on $\sim100$~day time scale. A flattening of the light curve, which is expected to accompany the transition (Waxman, Kulkarni \& Frail 1998, FWK00, Livio \& Waxman 2000), has been observed on a similar time scale in the radio light curves of most well-sampled afterglows \cite{Frail03}. 

Based on the radio observations of GRB970508 and on the hints for an association of GRBs with supernovae, Paczy\'nski (2001) suggested to search for radio emission from $\sim1$~yr old "GRB remnants" among nearby  ($<100$~Mpc) supernovae. Levinson et al. (2002) have shown that a large number of such remnants may be identified by all sky radio surveys. A radio survey monitoring 33 type Ib/c supernovae for $\approx1$~yr has recently been published by Berger et al. (2003b). The bright radio signature expected from (decelerated) GRB jets has not been detected, leading to the conclusion that the vast majority of type Ib/c SNe are not associated with cosmological ($L_{\gamma,\rm Iso.}\simeq10^{52}{\rm erg/s}$) GRBs. The fact that the radio luminosity of SN1998bw after $\sim1$~yr delay was 3 orders of magnitude lower than the expected $L_\nu\sim10^{30}\nu_{10\rm GHz}^{-1/2}{\rm erg/s\,Hz}$, appears therefore to rule out the "off-axis jet" interpretation of the low luminosity of GRB980425. We show here that this is not necessarily the case. 

The luminosity and spectrum of radiation emitted during jet transition to sub-relativistic expansion are determined by the total energy $E$ carried by relativistic plasma, by the number density $n$ of surrounding gas, by the fraction $\epsilon_B$ ($\epsilon_e$) of shock thermal energy carried by magnetic field (relativistic electrons), and by the shape of the electron distribution function, which is commonly assumed to be a power law of index $p\equiv{\rm d}\ln n_e/{\rm d}\ln \gamma_e\approx2$, where $\gamma_e$ is the electron Lorentz factor (FWK00). It is natural to assume that the values of $\epsilon_B$, $\epsilon_e$, and $p$ are  universal, since they are determined by the microphysics of the collisionless shock driven into the surrounding gas. Indeed, the constancy of $p$ and $\epsilon_e$ among different bursts is strongly supported by observations. In bursts where $p$ can be determined accurately (e.g. Galama et al. 1998a, FWK00, Stanek et al. 1999) $p=2.2\pm0.1$ is inferred, a value consistent with numeric and analytic calculations of particle acceleration via the first order Fermi mechanism in relativistic shocks \cite{RelShock1,Kirk00,RelShock2}. Universal values of $p$ and $\epsilon_e$ are also inferred from the clustering of explosion energies \cite{Frail01} and of X-ray afterglow luminosity\footnote{Apparently deviant values of $p$ \cite{CL99,PK02} are inferred based on light curves, rather than spectra, and are sensitive to model assumptions (e.g. they depend on the assumed radial dependence of the ambient medium density).} \cite{Freedman01,Berger03a}. The value of $\epsilon_B$ is less well constrained by observations. However, in cases where $\epsilon_B$ can be reliably constrained by multi waveband spectra, values close to equipartition are inferred (e.g. FWK00).

The total energy $E$ also appears to be universal. Although the apparent isotropic $\gamma$-ray energy, $E_{\gamma,\rm Iso.}$ varies by $\sim2.5$ orders of magnitude between different bursts, a strong correlation between $E_{\gamma,\rm Iso.}$ and $\theta_j$ is observed, implying a narrow distribution of beaming corrected $\gamma$-ray emission, $E_\gamma\equiv\theta_j^2 E_{\gamma,\rm Iso.}/2\approx10^{51}$~erg with roughly factor 3 spread \cite{Frail01}. Moreover, the beaming corrected X-ray afterglow luminosity (at fixed time), $L_X\equiv\theta_j^2 L_{X,\rm Iso.}/2$, which provides a robust estimate of the total kinetic energy carried by the relativistic plasma \cite{Freedman01}, is also approximately constant \cite{Berger03a}. Thus, variations in the radio flux during transition to sub-relativistic expansion are most likely due to variations in $n$, and possibly due to variations in $\epsilon_B$.

The prediction of $L_\nu\sim10^{30}\nu_{10\rm GHz}^{-1/2}{\rm erg/s\,Hz}$ at $\sim1$~yr delay was derived under the assumption of expansion into a uniform density medium with $n\sim1{\rm cm}^{-3}$, consistent with the inferred values of $n$, which typically lie in the range of $10^{0.5\pm1}{\rm cm}^{-3}$ (e.g.\\ \cite{Bloom03n}). Similar predictions hold, however, for expansion into a wind with\\ $\dot{m}\equiv(\dot{M}/10^{-5}M_\odot{\rm yr}^{-1})/(v_{\rm w}/10^3{\rm km\,s^{-1}})\simeq1$ \cite{LnW00}. $\dot{m}\sim1$ is typically adopted for modelling GRB afterglows in the scenario of expansion into wind (e.g. \cite{CL99}), since the massive stars believed to be the progenitors of SNe Ib/c associated with GRBs (e.g. \cite{Iwamoto98,Woosley99}), are observed to have winds with $\dot{m}\gtrsim1$ \cite{Willis91}. For the case of SN1998bw, the density of gas surrounding the progenitor is constrained by the observed radio emission, which is consistent with synchrotron emission from a supernova shock wave propagating into a wind \cite{WL99,Li99}. 

In order to address the question of whether or not an off-axis jet interpretation of GRB980425 is consistent with the long term radio observations of SN1998bw, we generalize in \S\ref{sec:model} the analysis of FWK00, which applies to expansion into a uniform medium, to the case of expansion into a wind. Our results are of general interest, beyond the analysis of GRB980425, since jet propagation into a wind, rather than into a homogeneous medium, may of course be characteristic for jets associated with SNe Ib/c in general. In \S\ref{sec:98bw} we discuss the implications of the results of \S\ref{sec:model} to the case of SN1998bw/GRB980425. Our conclusions regarding the nature of SN1998bw/GRB980425 and regarding the search for off-axis GRB signatures in nearby type Ib/c supernovae emission are summarized in \S\ref{sec:discussion}.

\section{Model}
\label{sec:model}

We first discuss in \S~\ref{sec:dynamics} the hydrodynamics. 
Synchrotron emission is discussed in \S~\ref{sec:synch}. Some aspects of the model applicability are discussed in \S~\ref{sec:applicability}.

\subsection{Dynamics}
\label{sec:dynamics}

Let us consider a conical jet of opening angle $\theta_j$ expanding into a $r^{-2}$ density profile created by stellar mass loss. We assume a density profile
\begin{equation}\label{eq:rho}
    \rho=K r^{-2},\quad K\equiv\frac{\dot{M}}{4\pi v_{\rm w}},
\end{equation}
where $\dot{M}$ is the mass loss rate and $v_{\rm w}$ is the wind velocity. As long as $\Gamma\gg1/\theta_j$, the transverse size of causally connected regions within the flow, $r/\Gamma(r)$, is smaller than the transverse size of the jet, $\theta_j r$, and the jet evolves therefore as if it were a conical section of a spherical relativistic blast wave. At this stage, the flow may be well described by the Blandford-McKee (1976) self-similar solutions, where
\begin{equation}\label{eq:Gamma}
    \Gamma=\left(\frac{9}{8\pi}\frac{E_{\rm Iso.}}{Krc^2}\right)^{1/2}.
\end{equation}
$E_{\rm Iso.}$ is the energy the flow would have carried had it been spherically symmetric. The true energy carried by the relativistic plasma is given by $E=\theta_j^2 E_{\rm Iso.}/2$ (assuming a two-sided jet). At a radius $r=R_\theta$ where $\Gamma(r)$ drops to $1/\theta_j$, $\Gamma(r=R_\theta)=1/\theta_j$, the transverse size of causally connected regions exceeds $\theta_j r$ and the jet starts expanding sideways. At this stage the jet Lorentz factor drops exponentially with $r$ \cite{Rhoads99}, and on a time-scale $\simeq R_\theta/c$ the flow approaches spherical symmetry. Using eq.~\ref{eq:Gamma} we have
\begin{equation}\label{eq:R_theta}
    R_\theta=\frac{9}{8\pi}\frac{\theta_j^2E_{\rm Iso.}}{Kc^2}=\frac{9Ev_w}{\dot{M}c^2}=
    1.5\times10^{18}\frac{E_{51}}{\dot{m}}\,{\rm cm},
\end{equation}
and the transition to spherical expansion takes place over a time scale
\begin{equation}\label{eq:t_theta}
    t_\theta\approx R_\theta/c=1.7\frac{E_{51}}{\dot{m}}\,{\rm yr}.
\end{equation}
Here, $\dot{m}\equiv(\dot{M}/10^{-5}M_\odot{\rm yr}^{-1})/(v_{\rm w}/10^3{\rm km\,s^{-1}})$. 

Eq.~\ref{eq:R_theta} implies that the mass enclosed within $r<R_\theta$ is comparable to $E/c^2$, and therefore that as the flow approaches spherical symmetry it also becomes sub-relativistic. For shock radii $R>R_\theta$ the flow approaches therefore the self-similar Sedov-von Neumann-Taylor solutions describing expansion of a spherical strong shock wave into a $r^{-2}$ density profile (e.g. Chapter XII of Zel'dovich \& Raizer 2002). In these solutions, the shock radius is given by
\begin{equation}\label{eq:R}
    R=\xi(\hat{\gamma})\left(\frac{E}{K}t^2\right)^{1/3}.
\end{equation}
Here $\hat{\gamma}$ is the adiabatic index of the gas, and $\xi(\hat{\gamma})$ is a dimensionless parameter of order unity. Exact determination of $\xi(\hat{\gamma})$ requires a numerical solution of the flow variable profiles. However, a straight forward generalization of the Chernyi-Kompaneets approximation for the Sedov-von Neumann-Taylor solutions in a homogeneous medium (e.g. Chapter I of Zel'dovich \& Raizer 2002) to the case of a $r^{-2}$ density profile yields an analytic approximation for $\xi$,
\begin{equation}\label{eq:xi}
    \xi(\hat{\gamma})=
      \frac{3}{2}\left[\frac{(\hat{\gamma}+1)^2(\hat{\gamma}-1)}{2\pi(7\hat{\gamma}-5)}\right]^{1/3}.
\end{equation}
For $\hat{\gamma}=5/3$, appropriate for sub-relativistic flow, we have $\xi=0.73$. Denoting by $R_{\rm SNT}$ the radius at which the shock velocity equals $c$, we have 
\begin{equation}\label{eq:R_SNT}
    R_{\rm SNT}=\frac{16\pi}{9}\xi^3\frac{Ev_w}{\dot{M}c^2}=
    0.38\times10^{18}\frac{E_{51}}{\dot{m}}\,{\rm cm}.
\end{equation}
We define $t_{\rm SNT}$ as the time at which the Sedov-von Neumann-Taylor solution gives $\dot{R}=c$,
\begin{equation}\label{eq:t_SNT}
    t_{\rm SNT}=2R_{\rm SNT}/3c=0.27\frac{E_{51}}{\dot{m}}\,{\rm yr}.
\end{equation}
Comparing Eqs.~\ref{eq:R_theta} and \ref{eq:R_SNT}, we infer that the flow becomes sub-relativistic as it approaches spherical symmetry. 

The wind density at $r=R_\theta$ is
\begin{equation}\label{eq:n_theta}
    n_\theta=0.12 \frac{\dot{m}^3}{E_{51}^2}\,{\rm cm}^{-3}.
\end{equation}
Eq.~\ref{eq:n_theta} shows that for $\dot{m}\simeq1$, the density $n_\theta$ is similar to that typical for the inter-stellar medium. Thus, the predicted radio signatures for a jet expanding into a $\dot{m}\simeq1$ wind are similar to those for a jet expanding into a typical inter-stellar medium \cite{LnW00}. $R_\theta$ in the case of expansion into uniform medium of density $n=1n_0{\rm cm}^{-3}$ is given by \cite{LnW00}
\begin{equation}\label{eq:R_theta_n0}
    R_\theta=\left(\frac{17}{4\pi}\frac{E}{\rho c^2}\right)^{1/3}=
    0.97\times10^{18}\left(\frac{E_{51}}{n_0}\right)^{1/3}\,{\rm cm},
\end{equation}
and the corresponding time scales are \cite{LnW00}
\begin{equation}\label{eq:t_n0}
        t_\theta=1.0 \left(\frac{E_{51}}{n_0}\right)^{1/3}\,{\rm yr},\quad 
        t_{\rm SNT}=0.20 \left(\frac{E_{51}}{n_0}\right)^{1/3}\,{\rm yr}.
\end{equation}

\subsection{Synchrotron emission}
\label{sec:synch}

At times $t\gg t_{\rm SNT}$ the flow is well described by the spherical non-relativistic self-similar solution, where the shock radius is given by Eq.~\ref{eq:R}. The long term afterglow observed in GRB970508 (FWK00) and several other afterglows \cite{Frail03} can be explained as synchrotron emission from electrons accelerated by the collisionless shock to relativistic energy. We assume that a constant fraction $\epsilon_e$ ($\epsilon_B$) of the post-shock thermal energy is carried by relativistic electrons (magnetic field), and that the electron distribution function follows a power-law, ${\rm d}\ln n_e/{\rm d}\ln \gamma_e=p$ for $\gamma_e\ge\gamma_m$. For simplicity, we assume that the shocked plasma is concentrated into a thin shell behind the shock, $R/\eta\ll R,$ within which the plasma conditions are uniform (the Chernyi-Kompaneets approximation). This implies, in particular, $\eta=(\hat{\gamma}+1)/(\hat{\gamma}-1)$.

Under the above assumptions, the scaling of magnetic field amplitude $B$ and of $\gamma_m$ with time is given by (compare to Eq. A2 in FWK00)
\begin{equation}\label{eq:Bscaling}
    B\propto t^{-1},\quad \gamma_m\propto t^{-2/3}.
\end{equation}
The flux and spectrum of synchrotron emission is then given by Eqs.~A3 to A12 of appendix A of FWK00, with power-law indices of the $t$ dependence in Eqs.~$\{$~A6,A7,A8~$\}$, $\{11/10,1-3p/2,-3\}$, replaced with $\{11/6,-1-7p/6,-7/3\}$ (due to the difference in temporal scalings between Eq.~\ref{eq:Bscaling} and Eq.~A2 of FWK00). These Eqs. imply, in particular, that the characteristic synchrotron frequency of electrons with $\gamma_e=\gamma_m$ is given by
\begin{equation}\label{eq:nu_m}
    \nu_m=2.3(3\epsilon_e)^2(3\epsilon_B)^{1/2}\frac{\dot{m}^{3/2}}{E_{51}}
    \left(\frac{t}{t_{\rm SNT}}\right)^{-7/3}\,{\rm GHz},
\end{equation}
and that the specific luminosity at frequencies $\nu\gg\nu_m$ (and below the cooling frequency) is, for $p=2$,
\begin{equation}\label{eq:L_nu}
L_\nu=1.7\times10^{30}(3\epsilon_e)(3\epsilon_B)^{3/4}\frac{\dot{m}^{9/4}}{E_{51}^{1/2}}
\left(\frac{\nu}{10{\rm GHz}}\right)^{-1/2}
\left(\frac{t}{t_{\rm SNT}}\right)^{-3/2}\,{\rm erg/s\,Hz}.
\end{equation}

Here too, the luminosity predicted for the case of expansion into $\dot{m}=1$ wind is similar to that predicted for expansion into a uniform density medium \cite{LnW00},
\begin{equation}\label{eq:L_nu_n0}
L_\nu=6.6\times10^{30}(3\epsilon_e)(3\epsilon_B)^{3/4}n_0^{3/4}E_{51}
\left(\frac{\nu}{10{\rm GHz}}\right)^{-1/2}
\left(\frac{t}{t_{\rm SNT}}\right)^{-9/10}\,{\rm erg/s\,Hz}.
\end{equation}

\subsection{Applicability and robustness}
\label{sec:applicability}

The flow approaches spherical symmetry on a time scale $t_\theta$ (Eq.\ref{eq:t_theta}), and since $t_{\rm SNT}<t_\theta$ (Eq.~\ref{eq:t_SNT}) it becomes (mildly) sub-relativistic at $t_{\rm SNT}<t<t_\theta$. Thus, an off-axis observer lying on a line of sight which makes a large angle, $\theta_{\rm o.a.}\simeq1$~rad, with the jet axis, will detect a flux comparable to that given by the model, Eq.~\ref{eq:L_nu}, at a time $t_{\rm SNT}<t_{\rm o.a.}<t_\theta$,
\begin{equation}\label{eq:t_oa}
    t_{\rm o.a.}\approx1\frac{E_{51}}{\dot{m}}\,{\rm yr}.
\end{equation}
Observers located closer to the line of sight will detect a higher radio flux at earlier times, $t<t_{\rm o.a.}$. At later times, $t\gg t_{\rm SNT}$, the flow approaches self-similarity and Eq.~\ref{eq:L_nu} provides a progressively more accurate approximation to the observed flux. On the other hand, the exact shape of the light curve at earlier time, $t\le t_{\rm o.a.}$, is highly model dependent: The sideways expansion and the deceleration of the jet depend on the spatial distribution within the jet of the energy density and the Lorentz factor. These distributions are poorly constrained by current observations. Moreover, for given energy density and Lorentz factor distributions, an accurate calculation of jet expansion and deceleration can only be carried out numerically (e.g. \cite{Ayal01,Granot01}). 

Granot \& Loeb (2003), for example, have recently considered emission from a point source on the jet axis observed by an off-axis observer in the case of expansion into a uniform density medium, with a simplified deceleration model. For $E=10^{51}$~erg and $n=1{\rm cm}^{-3}$ they find that an observer at $\theta_{\rm o.a.}\simeq1$~rad detects a flux similar to that detected by an on-axis observer at a time $t_{\rm o.a.}=0.4$~yr. The specific luminosity they obtain at this time is $1.2\times10^{29}{\rm erg/s\,Hz}$ at 43~GHz, assuming $\{\epsilon_e=0.1,\epsilon_B=0.01,p=2.5\}$, which (using the dependence of $L_\nu$ on $\epsilon_e$, $\epsilon_B$ given by Eq.~{\ref{eq:L_nu}) corresponds to $8.0\times10^{30}{\rm erg/s\,Hz}$ at 10~GHz for $\epsilon_e=\epsilon_B=1/3$. These results are similar to the analytic predictions of Eqs.~\ref{eq:t_n0},~\ref{eq:L_nu_n0}. 

In the Granot \& Loeb (2003) analysis, the flux rises at $t\le t_{\rm o.a.}$ nearly like a step function, and reaches at $t$ somewhat larger than $t_{\rm o.a.}$ a maximum, which is an order of magnitude higher than the flux detected by an on-axis observer. This behavior is due (as also pointed out by the authors) to the simplified analysis, where only emission from a point source on the jet axis is considered. In a more realistic model, which includes emission from the entire jet and where sideways expansion and deceleration are calculated in more detail, the flux detected by a $\theta_{\rm o.a.}\simeq1$~rad observer rises gradually at $t<t_{\rm o.a.}$ and the peak flux is expected to be (a few times) lower (e.g. \cite{Granot02}).

\section{Implications to SN1998bw/GRB980425}
\label{sec:98bw}

For $\dot{m}\simeq1$, as expected for GRB SN progenitors (see discussion following Eq.~\ref{eq:n_theta}), an observer at large, $\theta_{\rm o.a.}\sim1$~rad, offset is expected to measure at $t_{\rm o.a.}\sim1$~yr (Eq.~\ref{eq:t_oa}) a specific luminosity $L_\nu\sim10^{30}{\rm erg/s Hz}$ at 10~GHz (Eq.~\ref{eq:L_nu}, assuming $\epsilon_e\sim\epsilon_B\sim1/3$). For a burst at cosmological distance this implies $f_\nu\sim0.1$~mJy. This prediction is similar to that obtained for expansion into a homogeneous medium with density typical to that of the inter-stellar medium, $n\sim0.1{\rm cm}^{-3}$ (Eq.~\ref{eq:L_nu_n0}). Specific luminosities (fluxes) consistent with this prediction have been measured for GRB970508 (FWK00) and for several other long duration radio afterglows \cite{Frail03}. 

Figure~\ref{fig:bw98} compares the observed radio flux of SN 1998bw with the synchrotron flux predicted by the model described in \S~\ref{sec:model}. The upper set of solid curves corresponds to $\dot{m}=1$ (the peak in the 1~GHz light curve is due to self-absorption). The radio luminosity of SN1998bw/GRB980425 at 1~yr delay is 2.5 orders of magnitude lower than this model prediction. An off axis jet with "standard" energy $E=10^{51}$~erg and "standard" $\epsilon_e\sim1/3$ expanding into a wind with $\dot{m}=1$ can therefore be ruled out under the assumption $\epsilon_B\sim1/3$. An off-axis jet interpretation of GRB980425 may be consistent with observations if it is assumed that either the wind of the progenitor of SN1998bw was atypical, with $\dot{m}\ll1$, or that the magnetic field energy density fraction has been atypically low, $\epsilon_B\ll1$ (or both). Lower values of $\dot{m}$ and of $\epsilon_B$ reduce the specific luminosity at the transition to sub-relativistic expansion, which is proportional to $L_\nu\propto\dot{m}^{3/4}\epsilon_B^{3/4}$ (Eqs.~\ref{eq:L_nu},\ref{eq:t_SNT}). A low value of $\dot{m}\ll1$ also delays the time at which the flow approaches sub-relativistic expansion (Eq.~\ref{eq:t_oa}), and hence the time at which the radio emission approaches that of Eq.~\ref{eq:L_nu} for an off-axis, $\theta_{\rm o.a.}\sim1$~rad, observer, to $\gg1$~yr. 

The radio emission of SN1998bw is consistent with synchrotron emission from electrons accelerated by the shock wave driven into the wind by the supernova ejecta \cite{WL99,CL99}. Radio observations can therefore be used to constrain $\dot{m}$. As usually is the case for radio SNe, the data are not sufficient for determining all model parameters. In particular, there is a degeneracy in model predictions, that can be tested by observations, between $\dot{m}$ and $\tilde{\epsilon}_B$, the fraction of the supernova post-shock thermal energy carried by magnetic field. Interestingly, the radio light curves imply that either $\dot{m}$ or $\tilde{\epsilon}_B$ are atypically low: Data are consistent, e.g., with near equipartition value of $\tilde{\epsilon}_B$ and $\dot{m}\sim0.1$ \cite{CL99}, and also with $\dot{m}\sim6$
and $\tilde{\epsilon}_B\sim10^{-6}$ \cite{WL99,CL99}. 

If we adopt a model with $\tilde{\epsilon}_B\ll1$ and $\dot{m}\gtrsim1$, we would need to assume $\epsilon_B\le10^{-4}\ll1$ in order to reconcile radio observations with the predictions of an off-axis jet. In this case, the radio flux of the decelerated jet becomes undetectable. If, on the other hand, we adopt a model with  $\tilde{\epsilon}_B$ near equipartition and $\dot{m}\sim0.1$, the off-axis jet interpretation would be consistent with observations for ${\epsilon}_B$ near equipartition. In this case, the transition to sub-relativistic expansion takes place on $\sim10$~yr time scale, and assuming that we are at $\sim1$~rad offset from the jet axis, the radio flux from the decelerated jet could not have been observed during the $\sim1$~yr observations of SN1998bw. This is illustrated in Fig.~\ref{fig:bw98} by the lower set of solid curves which describe model predictions for $\dot{m}=0.1$. 

Several clarification comments should be made at this point. The off-axis jet interpretation of GRB980425 typically assumes that our line of sight is a few degrees away from the edge of a sharp-edged conical jet
\cite{Yamazaki03,Granot02}. For a jet with sharp edges, the flux drops rapidly when the line of sight makes an angle with the jet axis which is larger by $\sim1/\Gamma\sim0.01$ then the jet opening angle, which constrains our line of sight to be within a few degrees of the jet edge. This interpretation would be inconsistent with observations for the case of $\epsilon_B\sim1/3$ and $\dot{m}\sim0.1$, since in this case jet deceleration would bring our line of sight into the jet's radiation "beaming cone", leading to strong radio emission on time scale $<1$~yr. Our line of sight must make a large angle with the jet axis in order to avoid observing the strong radio emission from the decelerated jet at $t<1$~yr. The observed $\gamma$-ray flux of GRB980425 can be explained in this case by assuming that the jet is not sharp-edged, but rather has "wings" that extend to $\sim1$~rad and produce the observed low $\gamma$-ray luminosity. Alternatively, Compton scattering of photons into our line of sight may allow a large off-axis orientation \cite{Eichler99}.

For $\dot{m}=0.1$, the time at which an off-axis, $\theta_{\rm o.a.}\sim1$~rad, observer detects a flux comparable to that predicted by Eq.~\ref{eq:L_nu} is $t_{\rm o.a.}\simeq10$~yr (Eq.~\ref{eq:t_oa}). At this delay, the predicted radio flux is a few mJy, and the predicted optical flux is a few $\mu$Jy ($m_V=22.5$), assuming $\epsilon_B\sim1/3$. What is the current, $t\simeq5$~yr, flux predicted by this model? As discussed in \S~\ref{sec:applicability}, the detailed time dependence of flux observed at $t<t_{\rm o.a.}$ is highly model dependent. In realistic jet models, however, the flux is not expected to vary strongly at $t\sim t_{\rm o.a.}$. Hence, a $\simeq1$~mJy 10~GHz flux and $m_V\simeq23$ optical flux are expected in this model.

\section{Discussion}
\label{sec:discussion}

Simple analytic expressions are given in \S~\ref{sec:model} for the specific luminosity emitted by a GRB fireball expanding into a wind, after it had decelerated to sub-relativistic speed. At this stage radiation is emitted roughly isotropically. An off-axis observer lying on a line of sight which makes a large angle, $\theta_{\rm o.a.}\simeq1$~rad, with the jet axis, is predicted to detect a flux comparable to that given by the model, Eq.~\ref{eq:L_nu}, at a time $t\simeq t_{\rm o.a.}=1E_{51}\dot{m}^{-1}\,{\rm yr}$, where $E=10^{51}E_{51}$~erg. The specific luminosity at 10~GHz at $t\simeq t_{\rm o.a.}$ is $L_\nu\approx 0.24\times10^{30}(3\epsilon_e)(3\epsilon_B)^{3/4}\dot{m}^{3/4}E_{51}{\rm erg/s\,Hz}$.
At later times the model provides a progressively more accurate approximation to the observed flux. The exact shape of the light curves at earlier time, $t\le t_{\rm o.a.}$, is highly model dependent. In particular, it depends on the (unknown) spatial jet structure. 

The low radio luminosity of SN1998bw, compared to that expected from a decelerated GRB jet at $\sim1$~yr delay, may be consistent with the off-axis jet interpretation of GRB980425 provided that either the magnetic field energy fraction is atypically low, $\epsilon_B\le10^{-4}$, or that the density of the wind surrounding the progenitor is lower than typically expected, $\dot{m}\equiv(\dot{M}/10^{-5}M_\odot{\rm yr}^{-1})/(v_{\rm w}/10^3{\rm km\,s^{-1}})\simeq0.1$. Lower values of $\dot{m}$ and of $\epsilon_B$ reduce the specific luminosity at the transition to sub-relativistic expansion. A low value of $\dot{m}\ll1$ further delays the time at which the flow approaches sub-relativistic expansion, as illustrated in Fig.~\ref{fig:bw98}. In the former scenario, $\epsilon_B\le10^{-4}$, the flux from the decelerated jet becomes undetectable. We consider this scenario less likely, however, since we expect $\epsilon_B$, which is determined by the shock micro-physics, to be similar for different bursts, and $\epsilon_B$ close to equipartition is inferred from radio observations of other bursts. The latter scenario, $\dot{m}\simeq0.1$, is consistent with the constraints imposed on $\dot{m}$ by the observed radio emission from the supernova shock driven into the wind. In this scenario, transition to sub-relativistic expansion is expected over $\sim10$~yr time scale. A $\simeq1$~mJy 10~GHz flux and $m_V\simeq23$ optical flux are expected in this model at present, and the angular source size is expected to be $\approx10$~mas.

Our analysis also demonstrates that in order to place robust constraints on the fraction of SNe Ib/c associated with cosmological GRBs, radio follow up of such SNe may need to be carried over a multi-year period, since for lower values of $\dot{m}$ strong radio emission may be detected by an off-axis observer only after several years (Additional motivation for
multi-year monitoring is the possibility to identify radio emission from "failed GRBs," with
low velocity and large baryon load jets (Totani 2003)). The values of $n$ inferred from afterglow observations typically lie in the range of $10^{0.5\pm1}{\rm cm}^{-3}$ (e.g. \cite{Bloom03n}). As explained in \S~\ref{sec:dynamics} and \S~\ref{sec:synch} (see also \cite{LnW00}), the predicted long-term radio signatures for a jet expanding into a $\dot{m}\simeq1$ wind are similar to those for a jet expanding into a typical inter-stellar medium $n\sim 1{\rm cm}^{-3}$. For $n\sim1{\rm cm}^{-3}$, or $\dot{m}\sim1$, strong radio emission should indeed be observed by an off axis observer at $t\simeq1$~yr (see Eqs.~\ref{eq:t_oa},~\ref{eq:t_n0}). However, observations of GRB980425/SN1988bw demonstrate that, at least in some fraction of the cases, strong radio emission from a decelerated jet would be detected by an off-axis observer only over a longer time scale. 

Finally, it should be kept in mind that the wind mass-loss rate from the progenitor of SN1998bw may have been time dependent. The wind density profile at distances $\lesssim1$~pc corresponds to the wind history over $\sim10^3$~yr preceding the explosion (for a wind speed of $10^3$~km/s). We have no direct information on the steadiness of the mass loss rate from massive stars so close to the end of their evolution. A higher mass loss rate at earlier times, closer to the typically expected $\dot{m}\sim1$, would lead to a higher radio flux than predicted in figure~\ref{fig:bw98} at late times.

\acknowledgements
I thank B. Paczy\'nski for discussions that initiated the study described in this manuscript. 
EW is partially supported by AEC grant and a Minerva
grant, and is an incumbent of the Beracha Career development
Chair.

\clearpage

\begin{figure}
\plotone{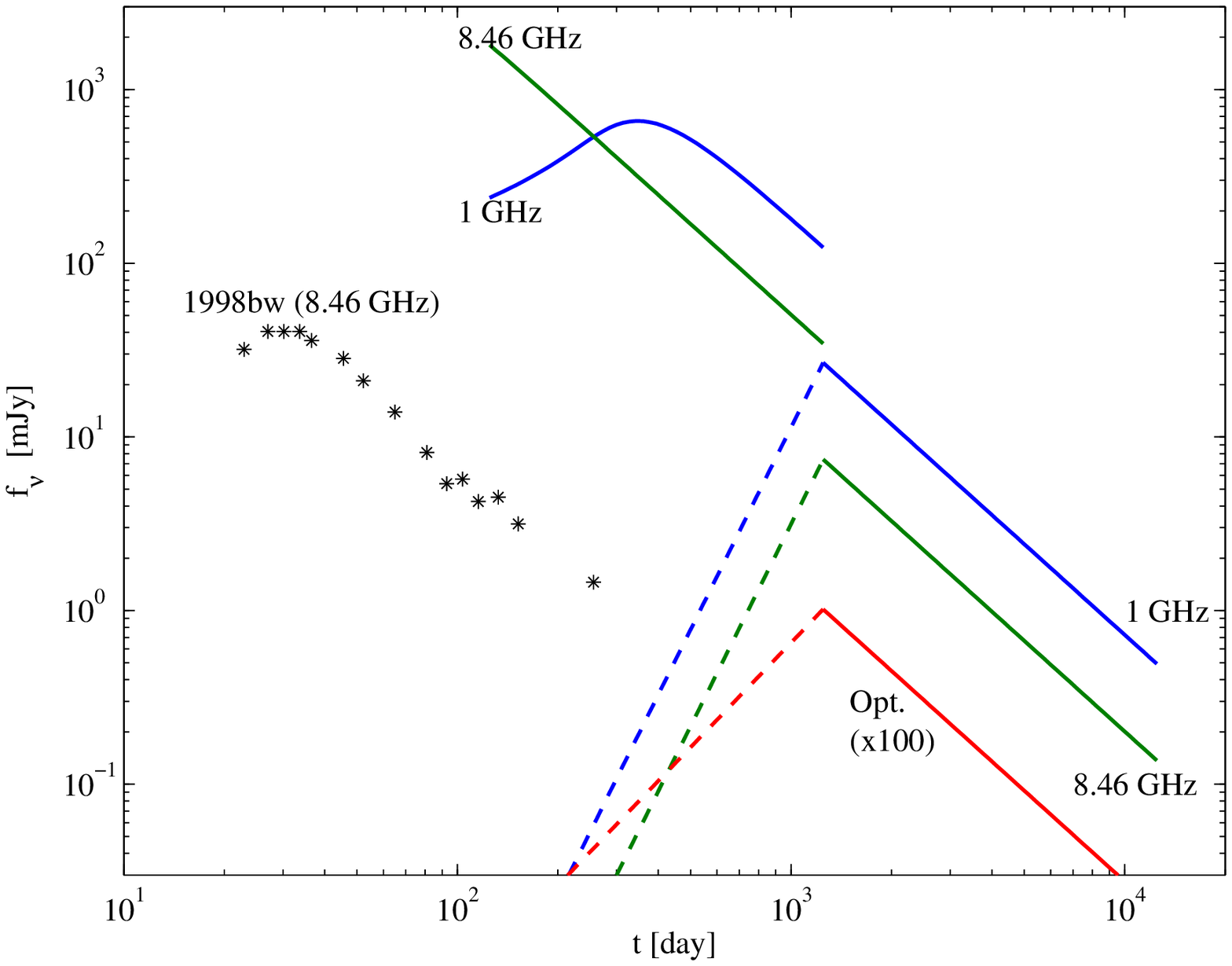}
\caption{The radio flux of SN 1998bw compared with predictions of the model described in \S~\ref{sec:model} for a jet of energy $E=10^{51}$~erg expanding into a wind. The upper, lower sets of solid curves show model flux at $t>t_{\rm SNT}$ for $\dot{m}=1$, $\dot{m}=0.1$ respectively. Equipartition, $\epsilon_e=\epsilon_B=1/3$, and $p=2.2$ are assumed. An off-axis observer lying on a line of sight which makes a large angle, $\theta_{\rm o.a.}\simeq1$~rad, with the jet axis, is predicted to detect a flux comparable to that given by the model, Eq.~\ref{eq:L_nu}, at a time $t\simeq t_{\rm o.a.}=1\dot{m}^{-1}\,{\rm yr}$. At later times the model provides a progressively more accurate approximation to the observed flux. On the other hand, the exact shape of the light curves at earlier time, $t\le t_{\rm o.a.}$, is highly model dependent. In particular, it depends on the (unknown) spatial jet structure. The dashed curves, added for illustration, follow the temporal dependence found in the numerical simulations described in Granot et al. (2002).
\label{fig:bw98}}
\end{figure}

\end{document}